\documentclass[sigconf]{acmart}
\usepackage[utf8]{inputenc}
\usepackage{scrextend}
\usepackage{blindtext}
\usepackage{indentfirst}

\begin{abstract}
    Blockchain and its distributed ledger technology have far-reaching implications for consumers across the world. Cryptocurrencies like XRP work to solve key issues in the remittance industry, targeting corridors like Mexico where foreign remittance fuels economies. Blockchain’s libertarian principles have the potential to change lives in the third world, replacing corrupt infrastructure with trust-based solutions. While this technology can be used to significantly improve lives, it has a wealth of destructive applications. Bitcoin’s blockchain and nefarious websites like the Silk Road have fueled an underground market of drugs, money laundering, and terrorism, complicating digital currency legislation. The negative environmental effects of cryptocurrency may also contribute significantly to global climate change. Negatives aside, cryptocurrency still proves to be a valuable commodity in technological development.
\end{abstract}

\begin{document}

\title{The Societal Implications of Blockchain Proliferation}
\author{Cory Cherven}
\affiliation{Canisius College \and Department of Computer Science}
\email{chervenc@canisius.edu}

\date{\today}
\copyrightyear{2020}
\setcopyright{acmcopyright}
\acmConference[CC '20]{The 150th Anniversary of Canisius College}{March 2020, Buffalo, NY}

\maketitle
\section*{Introduction}
The development of blockchain technology has created a new framework for doing immense societal good. A great benefit of blockchain is its effects of the remittance industry. Billions of dollars flow within and between countries every year, currently within a system designed and maintained by the Society of Worldwide Interbank Financial Telecommunication (SWIFT). This antiquated system acts as a tracking system for money as it flows through member banks (like post offices) but relies heavily on its member banks' infrastructures. As a result, it has proven to be immensely slow and transactions accumulate fees and delay at each step. Blockchain technology enables funds to be sent in real time with little delay, few fees, and without relying upon third parties to facilitate transfers. Cryptocurrencies like XRP \cite{ripplewhitepaper} are working to solve the antiquation of SWIFT's network by leveraging local cryptocurrency markets to instantaneously convert currency from the initial fiat, transfer it, and convert it to the destination fiat \cite{swift}.

These benefits of blockchain can also be felt in the third world where citizens lack trust in their governments and their banking systems. There are many intuitive applications for blockchain to aid these people such as: using micro-transactions to sell back solar energy to the grid and reduce costs \cite{africa}, storing money in decentralized wallets outside of the government's control \cite{mexico}, and as a means to store important and immutable identification records like IDs, medical records, etc. \cite{africa}. Blockchain, particularly its decentralized nature, can protect resources and information outside of the control of any single government, allowing third-world citizens to confidently protect themselves.

Unfortunately, blockchain's lack of centrality brings with it many detrimental effects on society as well. Bitcoin \cite{bitcoinwhitepaper} has proven to be a useful resource in online illicit marketplaces like the Silk Road which sell drugs, weapons, and many other nefarious goods and services \cite{silkroad}. By detaching identities and localities from digital bank accounts, cryptocurrencies on blockchains can evade money laundering regulation and government oversight \cite{laundering1}\cite{laundering2}. These currencies can be funneled to fund terrorism abroad while protecting the identities of donors and terrorist regimes \cite{terrorism}. Additionally, the immense electrical consumption of computers running blockchain software (particularly with proof-of-work blockchains, see glossary) contributes to rapid climate change and environmental degradation \cite{emissions}.  

Thankfully, many of the negatives of blockchain are counteracted by widespread government regulation. By tracing blockchain transactions at their entry points (currency exchanges), governments like the United States can follow the blockchain's serialized records to see where the money moves. By tagging identities to these exchange accounts, governments can effectively trace laundering and illicit transactions (including terrorism) \cite{perp}\cite{regulation}. While these regulations often undermine the original goal of Bitcoin (and many other cryptocurrencies) to be anonymous, they still enable blockchain technology's use for good in remittance and in aiding the unbanked. This regulation helps to ensure that blockchain technology does more good and society than harm and facilitates commercial and communal growth.


\section*{Background}
On January 3, 2009, a peer-to-peer (P2P) payment system known as Bitcoin went live  \cite{goodbad}. Its inventor, pseudonymously titled “Satoshi Nakamoto,” created the protocol to allow for P2P transfers without the need for a trusted third party (like a banking institution) in instigating a payment between two parties. Nakamoto \cite{bitcoinwhitepaper} envisioned a completely secure and anonymous means to transfer money that relied on computing power to verify a transaction’s legitimacy. The resulting token, Bitcoin, produced an upheaval in the way that data could move and be verified. Other digital currencies followed suit including Litecoin, Ripple, Dash, Dogecoin, and Peercoin \cite{regulation} which often reapplied Satoshi’s vision with tweaks, always utilizing similar technology. With the rise of the digital currency and Bitcoin’s climb to legitimacy in 2013 \cite{goodbad}, it became evident what positive and negative impacts the new technology would bring and that legislation was necessary.

There are many negatives that arise from an anonymous P2P network. Anonymity has spurred the growth of the “eBay of illicit drugs” \cite{silkroad}, allowed for money laundering \cite{laundering1}, funded terrorism \cite{goodbad}, and begun warming the earth \cite{emissions}. Thankfully, the traceability of blockchain and regulation of exchanges has helped counteract this anonymity and curbed crime.

Regulators of Bitcoin have avoided completely banning cryptocurrencies in most countries as they identify cryptocurrency’s benefits. Cryptocurrencies like Ripple (XRP) \cite{ripplewhitepaper} are working to overhaul legacy systems and reduce remittance fees to impoverished economies. Startups like BitPesa \cite{goodbad} are looking to reduce transaction fees and aid the unbanked in Africa. There remains an untapped potential to embed personal records (health, banking, identity) into blockchain to revolutionize banking, voting, and immigration issues that occur under unreliable governments  \cite{africa}.
\section*{Glossary}\noindent
\textbf{Node}–a computer hosting an instance of the Bitcoin protocol, responsible for verifying transactions. Anybody can host a node on the decentralized Bitcoin network.\\
\textbf{Block}–a chunk of data containing an assortment of information, particularly some or all the transactions since the last block was accepted by the network’s nodes.\\
\textbf{Difficulty}–how difficult it is to mine a block. This increases and decreases to ensure that a block is verified about every ten minutes.\\
\textbf{Mining}–the process by which the network verifies the transactions of a given block. All the computers (CPUs) on the network attempt to produce an SHA-256 hash of the block with a sufficient number of zeros at the beginning (this is determined by the difficulty) by iterating a nonce variable in the block and rehashing until it has enough zeros. By solving this hash, the winning CPU verifies the transactions on the block and sends the block to the other nodes to be accepted.\\
\textbf{Proof of Work (PoW)}–the process of creating the hash when mining, proves that the work has been done to verify the block, results in the receipt of the reward.\\
\textbf{Reward (incentive)}–each block mined has a special first transaction where the miner (or group of miners) is awarded a specific number of Bitcoins. This value halves about every four years and is currently 12.5 Bitcoins (February 2020) (this keeps the supply capped at 21,000,000 BTC).\\
\textbf{Consensus}–the process by which the network agrees on its state. After a block is mined, the network must verify that no coins have been double-spent and append that block to the blockchain. The network allows for several chains to form but the only honest chain (the one considered valid) is the longest one. Any errant chains caused by attackers will die off when consequent blocks cannot be added to it and it becomes too short.\\
\textbf{Public/Private Keys}–a Bitcoin owner can identify their Bitcoin "wallet" by a public key (used for exchanging the coins) which can be seen by anybody. They prove ownership of those coins by a private key in order to initiate a transaction. These keys are 256-bit numbers.\\
\textbf{Blockchain}–A living ledger consisting of all of the previously accepted blocks. A full blockchain can be traced back to the first transaction and shows each set of transactions necessary to get the network to its current state. The structure of the network makes it so that, once a transaction has occurred, it cannot be undone without recalculating every subsequent block. This results in a highly robust, secure system that could only be hacked if most nodes misbehaved and committed to verifying invalid blocks. This makes the Bitcoin network more secure as it gets more decentralized (with more nodes) \cite{bitcoinwhitepaper} \cite{goodbad}.\\
\textbf{Smart Contracts}–digital contracts that complete a specified action automatically when their terms are satisfied, often include payment after a number of steps are completed  \cite{mexico}.\\
\textbf{Money Laundering}–“the method by which all proceeds of crime are integrated into the banking systems and business environments of the world”  \cite{laundering1}.\\
\textbf{Remittance}–"The process of sending money to remove an obligation. This is most often done through an electronic network, wire transfer or mail" \cite{perp}.

\section*{Illicit Marketplaces}
In 2011, Bitcoin spurred the creation of one of the most sophisticated online illicit marketplaces hosted on the TOR network: Silk Road, the “eBay of illicit drugs”  \cite{silkroad}. Its creator, Ross Ulbricht, grew the website into a \$1.2 billion drug ring, hinging on the success of Bitcoin’s anonymous P2P transactions  \cite{goodbad}. Even after the website was brought down by the FBI in 2013, numerous websites sprouted up in its place to continue its goal  \cite{silkroad}. Fortunately, there still remain many traceable components to Silk Road’s successors, namely their use of the postal service to transport drugs. These flaws have resulted in a rise in drug seizures at ports of entry (particularly in Australia) and improved the ability for the DEA to track down dealers in the US  \cite{goodbad} \cite{silkroad}.

In 2012, Small \cite{goodbad} reports, a well-known and revered gynecologist, Dr. Olivia Bolles, utilized the anonymity of Bitcoin on the Silk Road to vend controlled substances as “MDPro.” She and her partner, Alexandra Gold, accumulated 610 transactions in a few months, catching the eye of the DEA. Bolles used a personal P.O. box as the return address on her sales, however, which led to the DEA tracing the drugs back to the source. A similar story unwound for the website’s founder Ross Ulbricht, or “Dead Pirate Roberts.” The FBI tricked Ulbricht into directing them toward a system administrator, Curtis Clark Green, for a buyer for bulk cocaine. When Ulbricht became aware that Green had been caught by the FBI, he hired an agent to kill Green–this spiraled into Ulbricht’s eventual capture in a public library on October 1, 2013.

These anecdotes suggest that, although Bitcoin’s anonymity appears to pose an unfair advantage for criminals in the drug trade, the native traceability provided by the public Bitcoin blockchain is potentially easier to trace than cash-only drug transactions \cite{goodbad}. Bitcoin continues to catalyze the market–many replacements for the Silk Road have since risen and fallen including Black Market Reloaded and Silk Road 2.0. There is, however, a benefit to the growth of internet cartels: reduced violence and safer product. Direct P2P drug transactions over the internet reduce the need for narco-traffickers and gangs in facilitating drug distribution. According to James Martin of Macquarie University \cite{silkroad}, this may curtail much of the violence associated with illicit drugs and, with more direct transactions, reduce costs and increase the purity of product, reducing the risk of overdoses on cutting agents like fentanyl.

Ultimately, drug distribution is not a new problem. The nature of Bitcoin has only digitized it, unintentionally offering a greater means for the federal government to control it. Martin \cite{silkroad} remarks that the rating system employed by the Silk Road improves product purity, protecting the customer from common cutting agents. The blockchain offers a valuable tool to trace the supply chain back to its source by following the transactions across the various blocks of the ledger. The success of the FBI in tracing Ulbricht (and Bolles, by the DEA) shows “that today's law enforcement agencies appear to have a firm grasp on how to conduct investigations and subsequent prosecutions of large-scale illegal online activities”  \cite{goodbad}.
\section*{Money Laundering}
From the anonymity capitalized upon by Ross Ulbricht comes the opportunity for another low-tech crime to evolve: money laundering. Small \cite{goodbad} reports that, upon its inception, 
\begin{addmargin}[1em]{2em}
	…a number of studies analyzing the block chain [sic] of the Bitcoin ledger demonstrated that there were users who were placing a large amount of money into one account, and over the course of a few hours, to a few days, transferring the money in small increments through hundreds of "dummy" accounts before it was recombined.
\end{addmargin}
Using online casino applications, laundering was also achieved at the time of Bitcoin’s inception. Online casinos that offer peer-to-peer poker games effectively allow a user to intentionally lose all their chips to another player to launder money into their hands  \cite{laundering1}. According to Stokes \cite{laundering1}, in 2012, anti-money laundering laws had not yet developed to accommodate digital currencies; as a result, cryptocurrencies posed a real risk of bypassing reporting requirements.

As of 2015, more modern methods of money laundering prevention have been launched which parallel traditional tracing mechanisms. Small \cite{goodbad} refers to Bitcoin much as the marked bills used to trace stolen fiat currencies. By tracing the Bitcoin through the public keys of users, governments can follow the currency into the hands of a publicly identifiable public key, such as that of a business. Once the Bitcoins reach a known party, agents can use that information to find the responsible party. Nakamoto was aware of this issue at Bitcoin’s inception, stating “[t]he risk is that if the owner of a key is revealed, linking could reveal other transactions that belonged to the same owner”  \cite{bitcoinwhitepaper}.

A simple approach to the laundering issue is regulation at the exchange level. Stokes \cite{laundering1} suggests that, while Bitcoin lacks identifying information within its ledger, if states impose regulations where fiat money enters and exits the system, they can target a great deal of laundering. As we will touch upon later, New York State’s BitLicense \cite{perp} is an introduction to exchange-side regulations that require customer information to be tied to Bitcoin's (and other cryptocurrencies') addresses. Thus, anytime laundered money touches a known account’s address (which is likely a large portion of them) it becomes trackable. There are some websites, however, that allow people to purchase cryptocurrency directly from individuals (like “localbitcoins.com”) which may still subvert this regulation.

Its traceability aside, there have been numerous cases thus far of money laundering utilizing cryptocurrency. According to Dyntu and Dykyi \cite{laundering2}, in May of 2013, a company entitled Liberty Reserve which maintained a system of electronic transactions laundered more than \$6 billion across the world. The company had millions of users which transferred money using their proprietary cryptocurrency the “Liberty Dollar (LD),” directing the payment back into USD at the endpoints. The Silk Road’s Bitcoin transactions were also completed using a type of money laundering obfuscation entitled “toggle-switch.” With it, all transactions were quasi-randomly directed to the intended user over many steps \cite{goodbad}.

Much like with the drug trade, the anonymity of Bitcoin and other cryptocurrencies has proven to be circumventable, as “[a]nyone can analyze the block chain [sic] and follow the trail of a singular bitcoin” \cite{goodbad}. Still, tools exist like “toggle switch” and Dark Wallet which mix transactions and complicate the tracking process \cite{laundering2}. Websites on the TOR network like Silk Road block IP information as well, another effort to make tracing laundering difficult on the Bitcoin network. Regulations like “BitLicense” aid in tracking and solving the issue of laundering, but it still remains quite possible with the use of various nefarious tactics on the blockchain and with direct bitcoin exchanges (on sites like “localbitcoins.com”).
\section*{Funding Terrorism}
Another significant risk that has arisen in the fight against terrorism is the use of cryptocurrencies to donate to terrorist organizations’ efforts. Bitcoin’s critics believe that its anonymity poses the possibility of circumventing counterterrorism efforts  \cite{goodbad}. Small \cite{goodbad} notes that efforts to track down Osama bin Laden succeeded by locating his personal courier. Had Osama bin Laden used Bitcoin as a means of funding, he would have needed fewer couriers to transfer physical money–thus, he may never have been located by the CIA. Often, mandatory reporting of suspicious monetary transfers provides governments with clues toward finding terrorist cells–Bitcoin could bypass these measures.

Much like the previous risks of Bitcoin and cryptocurrency, Bitcoin’s ability to fund terrorism stems from its anonymity. It, however, fails (in some regard) to replace current systems used by terrorist cells to transfer money known as “hawala” (“transfer” in Arabic) which operates as a black-market version of Western Union \cite{goodbad}. Hawala is natively less traceable than Bitcoin as there isn’t a public transaction ledger built into the system where law enforcement can sniff it out. There are currently good controls employed by the USA to counter hawala which include strong legal penalties worldwide for performing hawala transfers. Given this intrackability of hawala relative to Bitcoin, it makes less sense to employ Bitcoin’s system.

Still, with access to hawala in addition to cash couriers (like those employed by bin Laden), money transfer, and cryptocurrencies, the funding network for terrorism becomes far more extensive. Teichmann \cite{terrorism} notes some intrinsic issues with Bitcoin facilitating terrorism: it is difficult to transfer 1,000,000 euros without being spotted and destination markets could be illiquid and incapable to facilitate transfer to the local currency. The lack of banking regulations, however, works in terrorism’s favor where markets support it, and anonymity adds to the benefits. Cash is very bulky and often difficult to hide, hence why we often see photos in the media of large rooms full of laundered bills. Bitcoin is digital and as compact as the paper that a private key can be stored on, which is far easier to move across borders.

Bitcoin holistically has benefits in the underground markets used to support terrorism. Supporters that do not have access to (or that do not want to risk) hawala can easily transfer Bitcoin. Teichmann \cite{terrorism} notes that ransom fees are often best transacted in Bitcoin (much like ransomware employed on computers) and that Bitcoin is often directly convertible into weapons and munitions via the black market accessible by the TOR network (like the Silk Road). By using exclusively public networks and operating in countries with more relaxed regulation (like Slovakia, which has Bitcoin ATMs), terrorists can hide under governments’ radars while amassing wealth that hawala may not otherwise support. Teichmann \cite{terrorism} and Small \cite{goodbad} both acknowledge the capabilities of Bitcoin to support terrorism in some capacity.

\section*{Energy Consumption and Global Warming}
In 2018, Camila Mora et al. \cite{emissions} published a paper on the potential for Bitcoin emissions to contribute to global warming. Given that Bitcoin requires its miners to compute complex hashes at random, it requires a significant degree of computing power to verify transaction legitimacy. Mora et al. \cite{emissions} estimate that Bitcoin usage in 2017 emitted 69 million tons of carbon dioxide. Should Bitcoin adoption increase as forecasted, it is “capable of producing enough emissions to exceed 2°C of global warming in just a few decades”  \cite{emissions}. Unless the algorithms behind Bitcoin or the energy sources used are changed, this could have devastating consequences. Based on the best Bitcoin hardware available in 2016, it is estimated to consume 3.38 Terawatt hours of electricity to mine Bitcoin each year. To put that into context, in 2014, Jamaica’s 2.72 million people consumed 3.03 TWh of electricity \cite{socialism}.

Bitcoin’s infrastructure is reliant on a series of nodes that accumulate copies of the blockchain and millions of miners which verify the transactions on the blockchain. In order to maintain the network’s average mining rate of one block per ten minutes, the network increases the difficulty of the mining process  \cite{bitcoinwhitepaper}. As more miners jump onboard the Bitcoin network, it takes more computing power to mine the same number of Bitcoin, resulting in more electrical demand–this issue is also exacerbated by the halving of Bitcoin rewards for mining every four years  \cite{emissions}.

Although Bitcoin’s entire network was estimated to consume 73.12TWh in 2019, Zbinden and Kondova \cite{mexico} suggest that this is an acceptable expense in order to offset the energy used by the banking industry that Bitcoin strives to replace; this includes Credit cards, data centers, offices, vaults, etc. which result in about 100TWh of energy consumption annually. In addition to potential increases in block size (resulting in fewer blocks) as occurred with the Bitcoin Classic fork and increased energy efficiency, Zbinden and Kondova \cite{mexico} state that this energy use will only improve. Bitcoin, however has not succeeded in replacing the current banking infrastructure. At the current rate of Bitcoin adoption, Mora et al. \cite{emissions} estimate that Bitcoin will create an energy demand capable of producing enough emissions to raise global temperatures 2 degrees over the next few decades at the slowest rate of adoption or in as soon as 11 years.

\section*{Remittance}
Bitcoin and other cryptocurrencies are poised to make great, positive changes in the world of remittance, a \$400 billion industry. According to Tianyi Qiu et al., currently, money remittance is handled by the Society of Worldwide Interbank Financial Telecommunication (SWIFT). SWIFT’s interbank messaging system completes approximately 24 million transactions daily. The current system relies on partnerships with independent banks to complete the remittance, pushing transactions through five institutions before reaching their destinations–accumulating fees and delays as the money moves. Only the SWIFT messaging system (like a shipment tracker) is immediate \cite{swift}, the actual movement of funds is far slower. The cryptocurrency XRP \cite{ripplewhitepaper} has been working to solve the cost and delays of remittance by synchronizing the payment and messaging directly between two banks (or currencies)  \cite{swift}. Redefining remittance protocol is also on the docket for businesses like BitPesa, which, Small says, is targeting Africa (specifically Kenya) where remittance fees are far in excess of the norm (11\% vs 9\%, BitPesa is looking to charge 3\%)  \cite{goodbad}.

In Mexico, the remittance industry plays a large role in fueling local economies, especially after the manufacturing growth and migration of farmers resulting from the North Atlantic Free Trade Agreement (NAFTA). According to Zbinden and Kondova \cite{mexico}, the payments flowing into Mexico currently total \$28 billion, 10\% of Mexico’s total GDP growth rate. Average global remittance costs as of 2019 were 7\% via providers like Western Union. Thanks to the ubiquity of the Bitso and Volabit exchanges in Mexico and their cryptocurrency remittance applications, low-cost options are already arising.

Costs are not the only element of remittance being improved by cryptocurrency. The SWIFT infrastructure utilized by most banks for remittance is extremely antiquated and delayed. Qui et al. \cite{swift} note that, in some regions, money transfers using the SWIFT protocol can take up to five days. In a SWOT (strengths, weaknesses, opportunities, threats) analysis of the SWIFT messaging system, they show that SWIFT’s greatest issue is its lack of control over the money transferring process:
\begin{addmargin}[1em]{2em}
	…the real security and transaction time and cost for the payments to be settled might entirely depend on the sender banks, intermediate banks and receiving banks. The errors [sic] could occur within these processing banks which are out of control of SWIFT  \cite{swift}.
\end{addmargin}
This system, which has much room for error, could be greatly improved by involving crypto assets.

The cryptocurrency Ripple (XRP) \cite{ripplewhitepaper} proposes a means to overhaul or replace the SWIFT messaging system with a P2P system based upon distributed ledger technology (DLT). Using a public blockchain, much like Bitcoin, Ripple has designed several remittance products that source liquidity on cryptocurrency exchanges on either end of a transaction to transfer money. In doing so, they synchronize the messaging flow (transaction information, status, etc.) with the actual flow of value (the XRP currency) \cite{swift}. Within milliseconds, two banks are connected and a settlement either succeeds or fails between them; the banks are also provided with a wealth of information before the transaction such as estimated time, transaction fees, and exchange rates. Should there be different routes the settlement could take, multiple quotes will be provided, and the sending institution can choose the best rate available. As of 2018, more than 100 institutions and more than 75 countries have begun using Ripple technology for remittance–61 banks alone in Japan \cite{swift}. In November of 2019, a \$50 million investment made by Ripple into MoneyGram has secured the use of their technology in several corridors, particularly the USD-MXN corridor (remittance between the US and Mexico). This technology’s greatest threats, claim Qui et al. \cite{swift}, come from advancements in other cryptocurrencies, showing that blockchain will certainly develop into a valuable technology for remittance.
\section*{Aiding the Third World}
Africa and Mexico are two important testing grounds for Bitcoin and other cryptocurrencies beyond remittance. Africa is still developing into the modern world–since it lacks many infrastructures including electrical, banking, and the internet, new technologies are needed to compensate. Sean Button \cite{africa} suggests that Bitcoin’s microlending capacities (many small transactions) could aid with these deficits by allowing the sale of excess electricity back to grids and providing access to loans and credit. Both Africa and Mexico suffer from great political unrest and a general lack of trust toward banking institutions; Bitcoin allows for a store of value separate the government, one that is far more stable  \cite{mexico} \cite{africa}. Blockchain can immutably house identity documents for refugees and immigrants, host trustworthy voting, and organize shipping using smart contracts, allowing for better supply chain management  \cite{africa}.

Poverty in Mexico poses a great opportunity for blockchain improvement. Zbinden and Kondova \cite{mexico} cite that only 15 people own 5\% of Mexico’s wealth with 43.6\% of people living below the national poverty line. In 2005, less than 25\% of people in the capital used formal financial services (banks, financial institutions, etc.) due to high fees, distrust, and a lack of documentation. Although 63\% of Mexico City’s unbanked residents own their houses, they cannot legally prove ownership due to a lack of documentation. Government corruption is a great cause of this unbanking and public distrust in financial institutions.

Zbinden and Kondova \cite{mexico} report that Mexico, in 2017, has begun investing in “Blockchain HACKMX” which looks for blockchain solutions to corruption. The program has initiated a “smart” public tender which is a cashless smart-contract approach to popular votes on important government decisions–it will, however, be hosted in a private, government-controlled blockchain that could still be subject to manipulation. Mexico’s supportive Fintech (financial technology) regulations have spurred the proliferation of 140,000 Bitso ATMs which allow the unbanked to have Bitcoin wallets and complete currency remittance. This pro-Fintech, pro-cryptocurrency environment in Mexico has great potential to improve public trust in the government’s behavior.

In Africa, startups like BitHub Africa are pushing to launch cryptocurrency applications that aid in microlending, energy access, and crypto adoption. Button \cite{africa} notes that cryptocurrency loans detach individuals from untrustworthy central authorities in these corrupt countries and reduces administration fees. Cryptocurrencies provide a great solution for the 80\% of Sub-Saharan Africans that are unbanked, as supported by Bitcoin trading 33\% higher in Nigeria in 2017 than elsewhere in the world (greater confidence placed in cryptocurrency). The immutability of the blockchain poses a great opportunity to store identification records like biometrics, pictures, ID numbers, banking information, educational credentials, etc. This would allow documents to follow displaced citizens and reduce identity fraud. Voting, just like in Mexico, could be embedded into a blockchain whereby ID is verified, and votes are tallied immutably on the blockchain. Via smart contracts, shipping contracts can be closely tracked and completed as manufacturing burdens shift from China into Africa.

There is a great wealth of benefits that could come to African and Mexican (and other) societies as a result of the proliferation of blockchain technology. Such technology removes corrupt governments from arbitration and enables citizens to feel confident in their money, vote, and document security.
\section*{Legal Hurdles}
The biggest hurdles for blockchain development are legal ones. Sonderegger states that “Bitcoin is poised to threaten the very foundation upon which fiat currency and monetary policy rest: centralized control”  \cite{perp}. This scares any government whose economy is driven by a central bank–it takes away the government’s control. As mentioned, Bitcoin is often a haven for criminal activity and its decentralization makes it very difficult to influence. Fears of rapid deflation caused the by fixed Bitcoin supply lead experts to denounce its success as a currency  \cite{perp}. Many countries have continued to allow Bitcoin to flourish with limited restrictions by classifying its risks and taxation; at least 10 countries, however, have enacted complete (or near-complete) bans including China, Iceland, Russia, and numerous others  \cite{regulation}.

In 2013, the US Department of the Treasury’s Financial Crimes Enforcement Network (FinCEN) issued guidelines regarding virtual currency. According to Sonderegger \cite{perp}, Its guidelines imposed rules upon all entities creating, obtaining, exchanging, or otherwise dealing in digital currencies–establishing that these entities are subject to the rules governing normal money transmission. This began requiring that exchanges file with the federal government and comply with extremely complex state licensing. The 2014 case SEC v. Shavers ruled that Bitcoin is in fact (an investment of) money in the US, showing that Bitcoin could be considered a security though it does not perfectly fall into the bounds of the “Howey Test” used to determine security status. The Internal Revenue Service has agreed that, regardless of Bitcoin’s unclear status, it will be taxed as an investment. Many states have additionally instated plans like New York’s “BitLicense” which regulate any money laundering and cybersecurity risks inherent in cryptocurrency via close surveillance of the exchanges.

Internationally, responses to cryptocurrency have varied widely from monitoring to a total ban. According to Jan Lansky \cite{regulation}, only three countries have elected not to regulate (Croatia, Ireland, and Japan), yet eleven countries have completely banned them. Sonderegger \cite{perp} suggests, however, a middle-ground approach to regulation which mirrors the Chinese approach to cryptocurrencies where: (1) they are not a currency (rather, a commodity), (2) banks and financial institutions cannot transact using them, (3) exchanges are legal (with government oversight), and (4) the population must assume all of the risk. The only variation is that, unlike China, cryptocurrency should be allowed in the pricing and purchasing of goods and services. As discussed, however, barring financial institutions from using cryptocurrencies poses a big risk for currencies like Ripple (XRP) \cite{ripplewhitepaper} which plan to revolutionize the remittance industry and aid compromised governments’ citizens.
\section*{Conclusions}
Blockchain technology is poised to make great positive changes in society if legislatures are prepared to consider the challenges of a society where anonymous P2P transactions are the norm. Under the guise of complete anonymity, the potential for expansive online illicit marketplaces to emerge arises \cite{goodbad} \cite{silkroad}, money is more easily launderable \cite{laundering1}, terrorists can secure private funding \cite{terrorism}, and heavy mining costs drive carbon emissions and global warming  \cite{emissions}. While many of these threats of cryptocurrency can be counteracted by strong regulation and the traceability of blockchain technology, some risks are still inevitable byproducts of advancements (particularly carbon emissions, which are unlikely to slow). Thankfully, there are other consensus protocols, like Ripple's \cite{ripplewhitepaper} which do not depend on inefficient mining and have relatively low carbon emissions compared to proof-of-work consensus.

Governments should tolerate these pitfalls for the benefits in remittance \cite{swift} \cite{goodbad} and for the potential to improve corrupt governments \cite{mexico} \cite{africa} brought by blockchain technology. The technological implications of the unmodifiable blockchain provide many great opportunities for the unbanked, for citizens of corrupt governments, and for those refugees and migrants whose identities must be as mobile as themselves. There is still a lot of room to develop a legal standpoint on cryptocurrency which does not impede on its free growth, and plenty more debate surrounding topics of cybercrime (ransomware, botnetting, and the security of distributed ledger technology), cryptocurrencies as investment vehicles, and–especially–the benefits of cryptocurrency for first-world citizens (like Ripple’s Xpring initiative).

Though they come with complex consequences, digital currencies provide a great opportunity for growth across the world and, as a young technology, they remain vastly unexplored. Further development will certainly expose greater flaws and benefits to the technology and additional legislative challenges to be overcome.

\bibliographystyle{ACM-Reference-Format}
\bibliography{mybibliography}

\end{document}